\documentclass{amsart}
\usepackage{amsfonts}
\usepackage{amsmath}
\usepackage{amsmath}
\usepackage{amsaddr}
\usepackage{graphicx}
\begin{document}
\title{Sector-based Factor Models for Asset Returns}
\author{A. Gu}
\address{Phillips Andover Academy}
\author{P. Zeng}
\address{Princeton University}
\begin{abstract}
Factor analysis is a statistical technique employed to evaluate how observed
variables correlate through common factors and unique variables. While it is
often used to analyze price movement in the unstable stock market, it does not
always yield easily interpretable results. In this study, we develop improved
factor models by explicitly incorporating sector information on our studied
stocks. We add eleven sectors of stocks as defined by the IBES, represented by
respective sector-specific factors, to non-specific market factors to revise
the factor model. We then develop an expectation maximization (EM) algorithm
to compute our revised model with 15 years' worth of S\&P 500 stocks' daily
close prices. Our results in most sectors show that nearly all of these factor
components have the same sign, consistent with the intuitive idea that stocks
in the same sector tend to rise and fall in coordination over time. Results
obtained by the classic factor model, in contrast, had a homogeneous blend of
positive and negative components. We conclude that results produced by our
sector-based factor model are more interpretable than those produced by the
classic non-sector-based model for at least some stock sectors.

\end{abstract}
\maketitle

\section{Introduction\label{sec1}}

Suppose we visit a local high school and decide to study the academic
performances of a random sample of students. We model each student's GPA as a
random variable. We can expect these GPAs to vary widely. At the same time,
though, we can also expect these GPAs to somehow relate to each other through
common factors.

To understand how the GPAs vary, we can assume that only a very small number
of aspects of students' lives account for the majority of their GPAs. For how
many hours a day do they watch the television, play sports, or socialize with
friends? How long do they use their computers to read the news, check their
emails, use social networking sites, or finish homework? How much do they sleep?

Each of these aspects is a random variable, and, among the students, we would
expect a varied distribution. Yet we would suspect that, with a single
student, GPA and the measurements would not all be truly independent; some
underlying factors would connect and influence some of these measurements. For
example, it would not be a wild guess to assume that a student who spent nine
hours a day sprawled in front of the television but only five minutes in front
of a book had a lower GPA than average. Socialization could be related to time
spent finishing homework or checking email through a common factor.

Of course, students are individuals. They, like the real world, are so
infinitely complex that we cannot even dream of precisely finding and
calculating the effects of all the factors that influence GPA, no matter how
many factors we can think of and express statistically.

Factor analysis is meant to relate observed variables $X_{1},X_{2}%
,X_{3},\ldots,X_{n}$ with a small number of common factors and a special
random variable unique to each $X$ \cite{2,5,6,8,18,23}. We seek to explain
how the variables are interconnected by these common factors. The unique
random variables are included to account for the variability not influenced by
any of the common factors. We could perform factor analysis on the
hypothetical study mentioned above on high school students to see how factors
might impact students' academic performance.

The common factor model is as follows:%
\begin{equation}
X=\mu+\Lambda F+\varepsilon\label{eq1.1}%
\end{equation}
where $X$ and $\mu$ are $n\times1$ vectors; $\Lambda$ is an $n\times m$
matrix; $F$ is an $m\times1$ vector; and $\varepsilon$ is an $n\times1$
vector. It may be clearer to the reader if we expand equation:%
\begin{equation}%
\begin{bmatrix}
x_{1}\\
x_{2}\\
x_{3}\\
\vdots\\
x_{n-1}\\
x_{n}%
\end{bmatrix}
=%
\begin{bmatrix}
\mu_{1}\\
\mu_{2}\\
\mu_{3}\\
\vdots\\
\mu_{n-1}\\
\mu_{n}%
\end{bmatrix}
+%
\begin{bmatrix}
\Lambda_{11} & \Lambda_{12} & \cdots & \Lambda_{1m}\\
\Lambda_{21} & \Lambda_{22} & \cdots & \Lambda_{2m}\\
\Lambda_{31} & \Lambda_{32} & \cdots & \Lambda_{3m}\\
\vdots & \vdots & \ddots & \vdots\\
\Lambda_{n-1,1} & \Lambda_{n-1,2} & \cdots & \Lambda_{n-1,m}\\
\Lambda_{n1} & \Lambda_{n2} & \cdots & \Lambda_{nm}%
\end{bmatrix}%
\begin{bmatrix}
F_{1}\\
F_{2}\\
\vdots\\
F_{m}%
\end{bmatrix}
+%
\begin{bmatrix}
\varepsilon_{1}\\
\varepsilon_{2}\\
\varepsilon_{3}\\
\vdots\\
\varepsilon_{n-1}\\
\varepsilon_{n}%
\end{bmatrix}
. \label{eq1.2}%
\end{equation}

Thus each $X $ is represented as a linear combination of random variables $F$
(factors), with the components of $\Lambda$ as coefficients. In the
hypothetical academic performance study proposed above, $x_{j}$ is the current
GPA of student $j$, and $\mu_{j}$ is his ``true'' GPA over the course of the
study. $F_{1},F_{2},\dots,F$ are latent unobserved random variables meant to
interpret $x_{j}$ up to error $\varepsilon_{j}$. $\Lambda$ is the factor
loading matrix with $\Lambda_{jk}$ being the weight of factor $F_{k}$,
referred to as loading---it indicates the relative importance of $F_{k}$ to
$x_{j}$. A low value of $\Lambda_{jk}$, therefore, indicates that its $F_{k}$
does little in influencing $x_{j}$.

In order to successfully run factor analysis, we must make several
assumptions. First, all $\varepsilon_{j}$ and $F_{j}$ are normal with $0$ mean
and mutually independent. We also assume that $\varepsilon_{j}$ and $F_{j}$
have standard deviations of $d_{j}$, and $1$, respectively. This implies that
$x_{j}$ is also normal with $0$ mean. Factor analysis is easy to use and often
provides very useful approximations in practice.

The history of factor analysis is rooted in psychology. In 1904, Charles
Spearman published an article in the \textit{American Journal of Psychology},
trying to find a definitive and completely accurate measure of intelligence
\cite{21}. Factors that influenced intelligence, he argued, were the subject's
test scores on pitch, light, weight, classics, French, English, and
mathematics; the scores on those tests were weighted. From then on, factor
analysis has expanded from psychology to many different fields of study
\cite{1,2,3,4,7,8,9,13,15,18,20,22,24}, along with broad discussions about how
to efficiently compute factor models \cite{2,10,11,14,16,17,19,23}.

In particular, it is a well-established practice to use factor analysis to
study price fluctuations in stock prices \cite{2,6}. The stock market is
notable for its seeming randomness and frequent large up and down swings. By
studying the daily log return, i.e.,
\[
\log\frac{\text{stock price at the end of a day}}{\text{stock price at end of
previous day}},
\]
of stock prices through factor analysis, for example, we can better understand
these price swings and potentially design trading strategies that can better
handle them.

Nonetheless, factor analysis is not at all perfect and is quite ambiguous
\cite{1}. The factor loading matrix $\Lambda$ is far from unique. In the
hypothetical academic performance study example, there is no obvious
relationship between $\Lambda$ and any identifiable aspect of life that might
influence academic performance. The motivation for a factor model is not
actually built into the factor model. This paradox can often limit the
usefulness of factor models. There have been many attempts to improve the
factor model for better practical performance \cite{6,10}.

In the stock market, stocks are often divided into sectors depending on the
products of the associated companies. According to the Institutional Brokers'
Estimate System (IBES), we have used $11$ sectors: finance, health care,
consumer non-durables, consumer services, consumer durables, energy,
transportation, technology, basic industries, capital goods, and public
utilities. For example, Morgan Stanley (stock symbol: MS) is in the financial
services sector. Its stock price often moves with other financial services
stocks. Google (stock symbol: GOOG) provides internet-related services and is
in the technology sector. Its stock price often moves with other technology
stocks as well. There are also stocks that do not belong to any of these
sectors. When factor models are used to measure variability of stock price
fluctuations, such sector-dependent behavior is typically not revealed in the
factor model, although some sector association can be seen (see
Sections~\ref{sec3} and \ref{sec4}.)

It is the goal of our work to develop factor models that explicitly utilize
sector information, to better exploit the real relationships between stocks.
We first select our factor-loading matrix to ensure representation of both the
whole market and every sector. We assign each sector to a different $F_{k}$:
$F_{1}$ represents finance, $F_{2}$ represents health care, and so on. We
choose $m>11$, such that $F_{12}$ through $F_{m}$ are non-specific market
factors. We set $\Lambda_{jk}$ to be zero when $x_{j}$ is not in the industry
represented by $F_{k}$. For example, with $m=13$, a business $x$ in the
transportation industry ($F_{7}$) would be expressed as (compare with equation
\eqref{eq1.2})%
\[
x=\mu_{j}+\Lambda_{j1}F_{1}+\Lambda_{j2}F_{2}+\dots+\Lambda_{jm}%
F_{m}+\varepsilon_{j},
\]
or%
\[
x=\Lambda_{j7}F_{7}+\Lambda_{j,12}F_{12}+\Lambda_{j,13}F_{13}+\varepsilon
_{7}.
\]

We then develop an expectation maximization (EM) algorithm to compute the
factor model, and use this algorithm to test our factor model using daily
close prices of the S\&P 500 stocks over the last 15 years. Finally, we drew
some interesting conclusions from our tests.

\section{Materials and Methods\label{sec2}}

In the study of daily stock log returns, we will make the typical assumption
that stocks have $0$ returns over time so that $\mu=0$ in equation
\eqref{eq1.1}. Our goal is to develop a method to identify $\Lambda$ and
$\Psi$, the diagonal covariance matrix of $\varepsilon$, with sector
information specifically built into $\Lambda$ and by observing $X  $ over
a period of time.

Equation \eqref{eq1.1} allows us to write out the covariance matrix for $X$
explicitly as $\Lambda\Lambda^{T}+\Psi$. To proceed, we first review a simple
method to identify $\Lambda$ and $\Psi$ in the standard factor model, which
repeatedly improves the estimates for $\Lambda$ and $\Psi$ based on the
principle of expectation maximization (EM). More details are in \cite{10,17}.

Given a pair of estimated $\Lambda$ and $\Psi$, the expected value of the
factors is%
\[
E(F\mid X)=\beta X,
\]
where $\beta=\Lambda^{T}(\Psi+\Lambda\Lambda^{T})^{-1}$, and the second moment
of the factors is
\[
E(FF^{T}\mid X)=I-\beta\Lambda+\beta XX^{T}\beta^{T}.
\]

Given daily log returns $X_{1},X_{2},X_{3},\dots,X_{p}$, we then compute a
pair of new estimates for $\Lambda$ and $\Psi$ through an Expectation step
(\textbf{E}-step) and an Maximization step (\textbf{M}-step):

\begin{description}
\item[\textbf{E}-step] Compute $E(F\mid X_{i})$ and $E(FF^{T}\mid X_{i})$ for
all data points $i$.

\item[\textbf{M}-step]
\begin{align*}
\Lambda^{\text{new}}  &  =AB^{-1},\\
\Psi^{\text{new}}  &  =\frac{1}{p}\operatorname{diag}\left(  \sum
\nolimits_{i=1}^{p}X_{i}(E(F\mid X_{i})^{T}-\Lambda^{\text{new}}%
\sum\nolimits_{i=1}^{p}E(F\mid X_{i})X_{i}^{T}\right)  ,
\end{align*}

\end{description}

where $A=\sum_{i=1}^{p}X_{i}E(F\mid X_{i})^{T}$ and $B=\sum_{i=1}^{p}%
E(FF^{T}\mid X\_i)$.

To build the IBES sector information into the factor model, we will choose
$m\geq12$ so that there are $11$ sector factors and $m-11$ market factors in
the factor model. The market factors reflect the dependence of individual
stocks on a few number of global market parameters, but the sector factors
will only reflect inter-dependence of stocks within a given sector. As
mentioned in Section~\ref{sec1}, we choose to set $\Lambda_{jk}$ to $0$ if
stock $j$ is not in sector $k$. This way, there are quite a lot of zeros in
the factor-loading matrix $\Lambda$, reflecting on the financial intuition
that stocks in different sectors have less influence over each other (although
they can still exert influence through the global market factors).

Remarkably, only small modifications need to be made to identify this new
factor model. For stock $j$, let $\mathcal{I}_{j}$ be the set of non-zero
loading indexes. Then,%

\[
\Lambda^{\text{new}}(j,\mathcal{I}_{j})=A(j,\mathcal{I}_{j})\left(
B(\mathcal{I}_{j},\mathcal{I}_{j})\right)  ^{-1}%
\]

The formula for $\Psi^{\text{new}}$ remains the same. See details in
Appendix~\ref{appB}.

\section{Results and Discussion\label{sec3}}

To validate our factor model, we downloaded daily close prices of the S\&P 500
stocks over 15 years (from 1996 to 2010). There were actually $854$ stocks
over this time period due to the frequent changes in the index. We also
obtained IBES sector classifications over this same period.

Our mentor provided a MATLAB program, which we used to perform an empirical
study of the new factor model on the daily returns computed from these close
prices. For this experiment, we choose $m=13$ and convergence is considered
reached after $100$ EM iterations.

\begin{figure}[!htbp]
\label{fig41}
\caption{}
\includegraphics[height=3.5in]{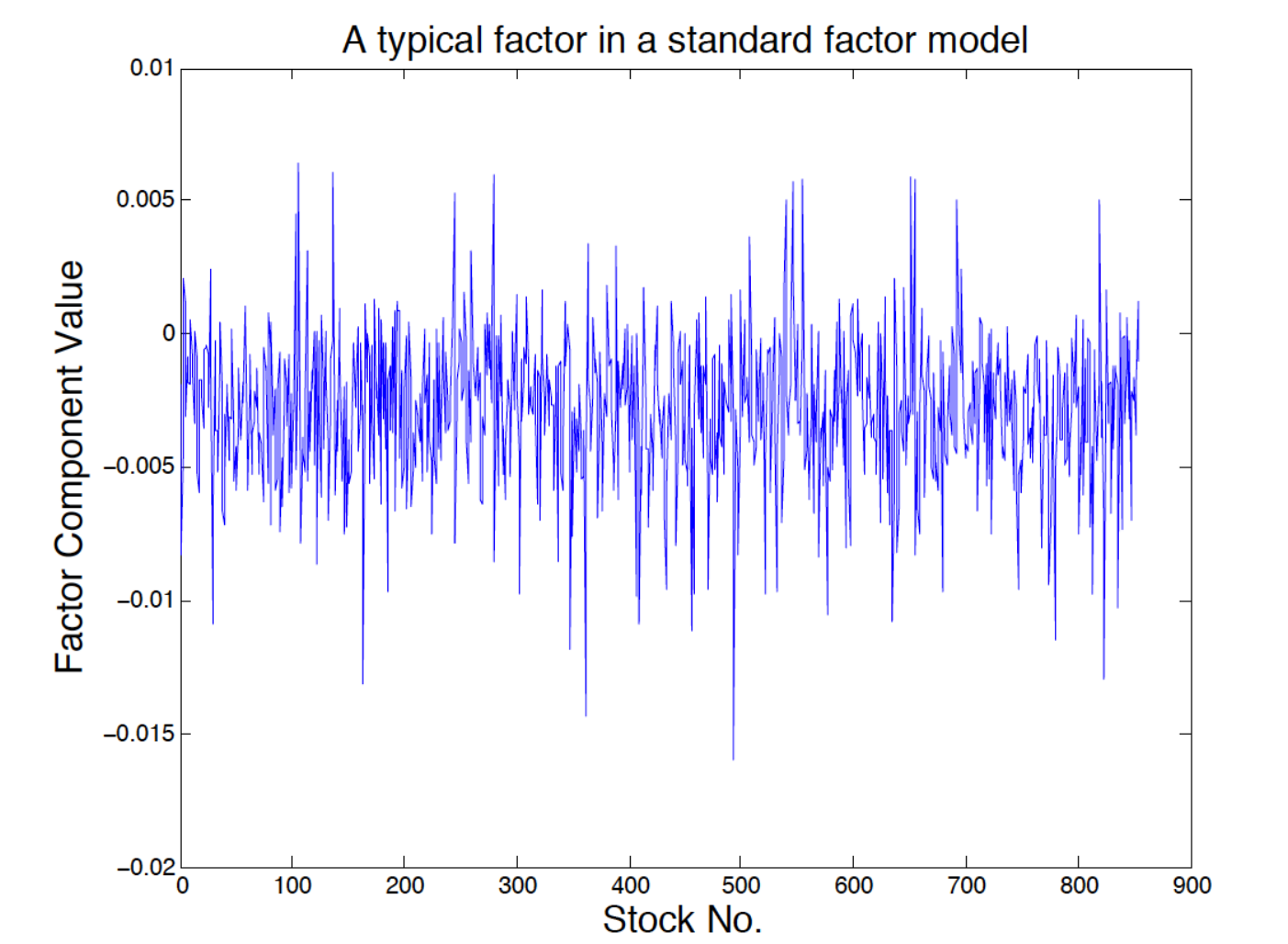}
\end{figure}

\begin{figure}[!htbp]
\label{fig42}
\caption{}
\includegraphics[height=3.5in]{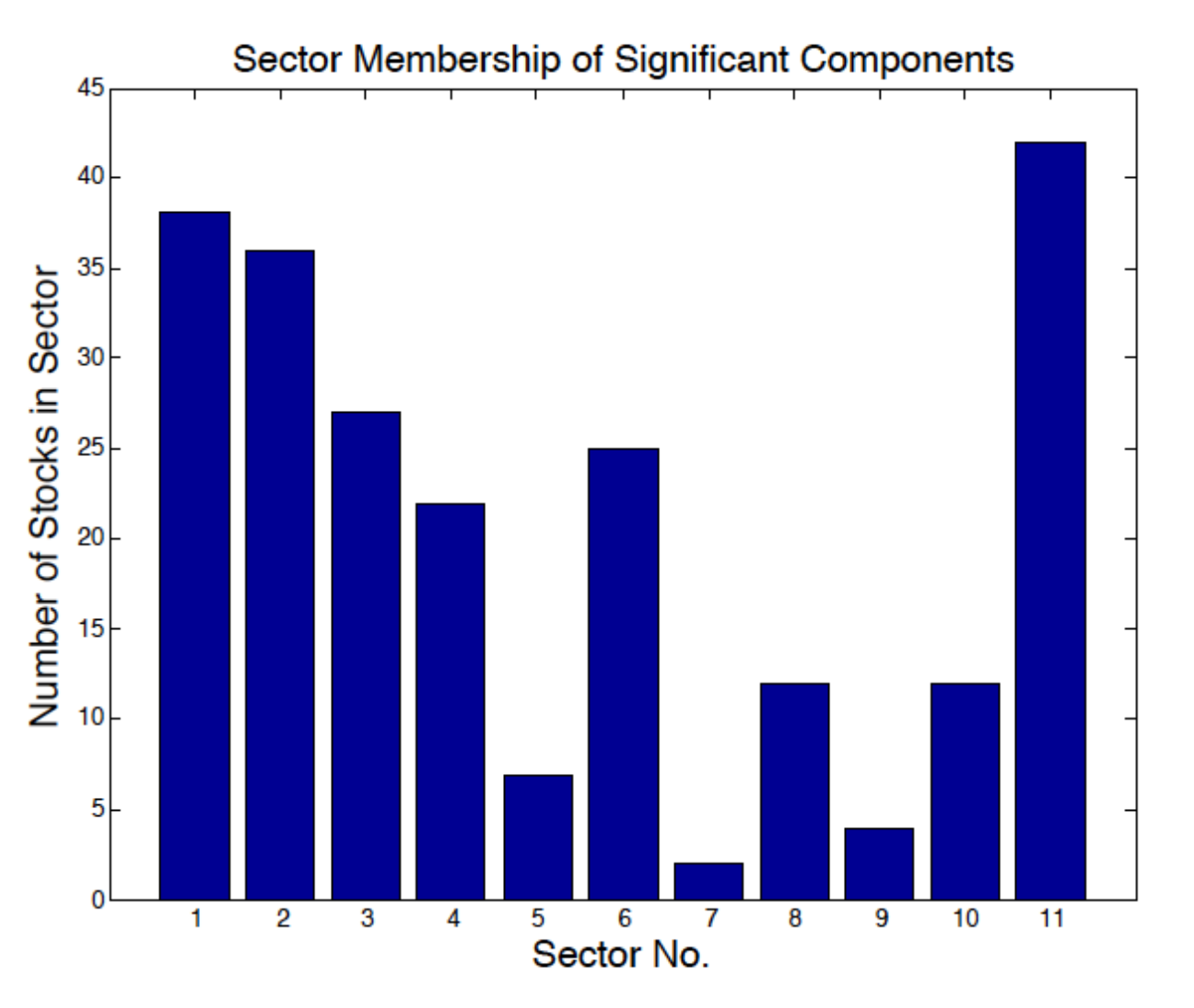}
\end{figure}

\begin{figure}[!htbp]
\label{fig43}
\caption{}
\includegraphics[height=3.5in]{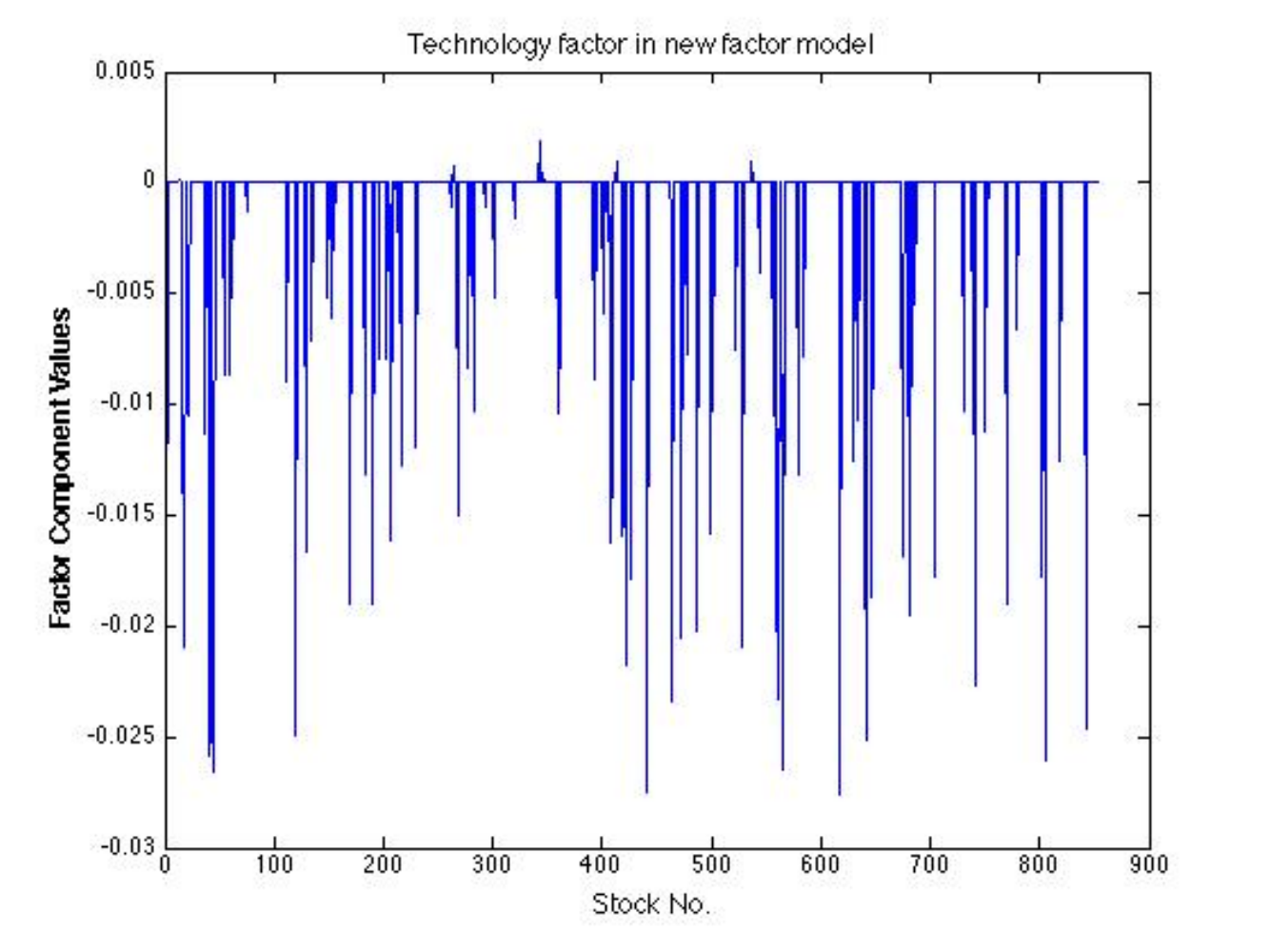}
\end{figure}

\begin{figure}[!htbp]
\label{fig44}
\caption{}
\includegraphics[height=3.5in]{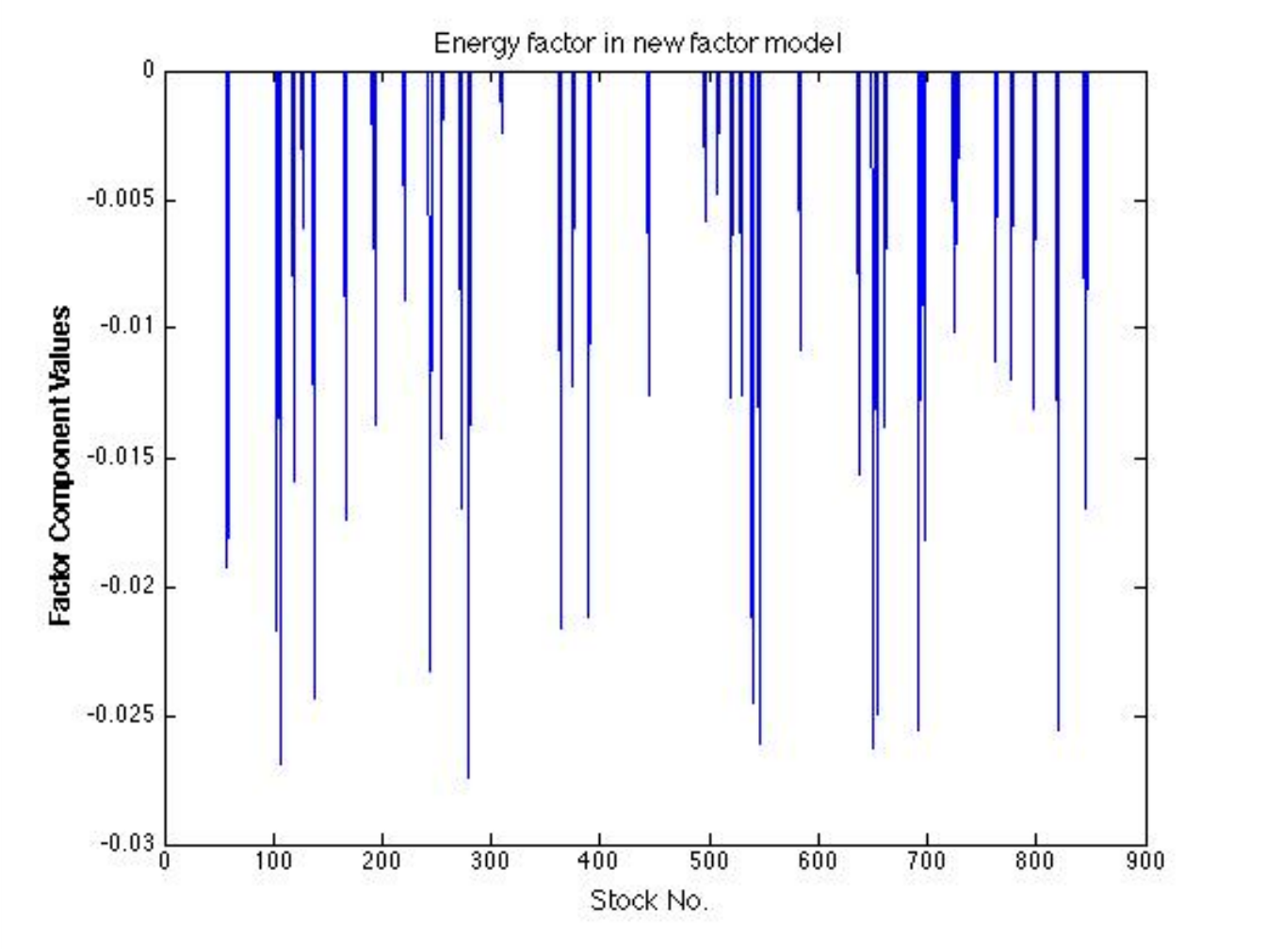}
\end{figure}

\begin{figure}[!htbp]
\label{fig45}
\caption{}
\includegraphics[height=3.5in]{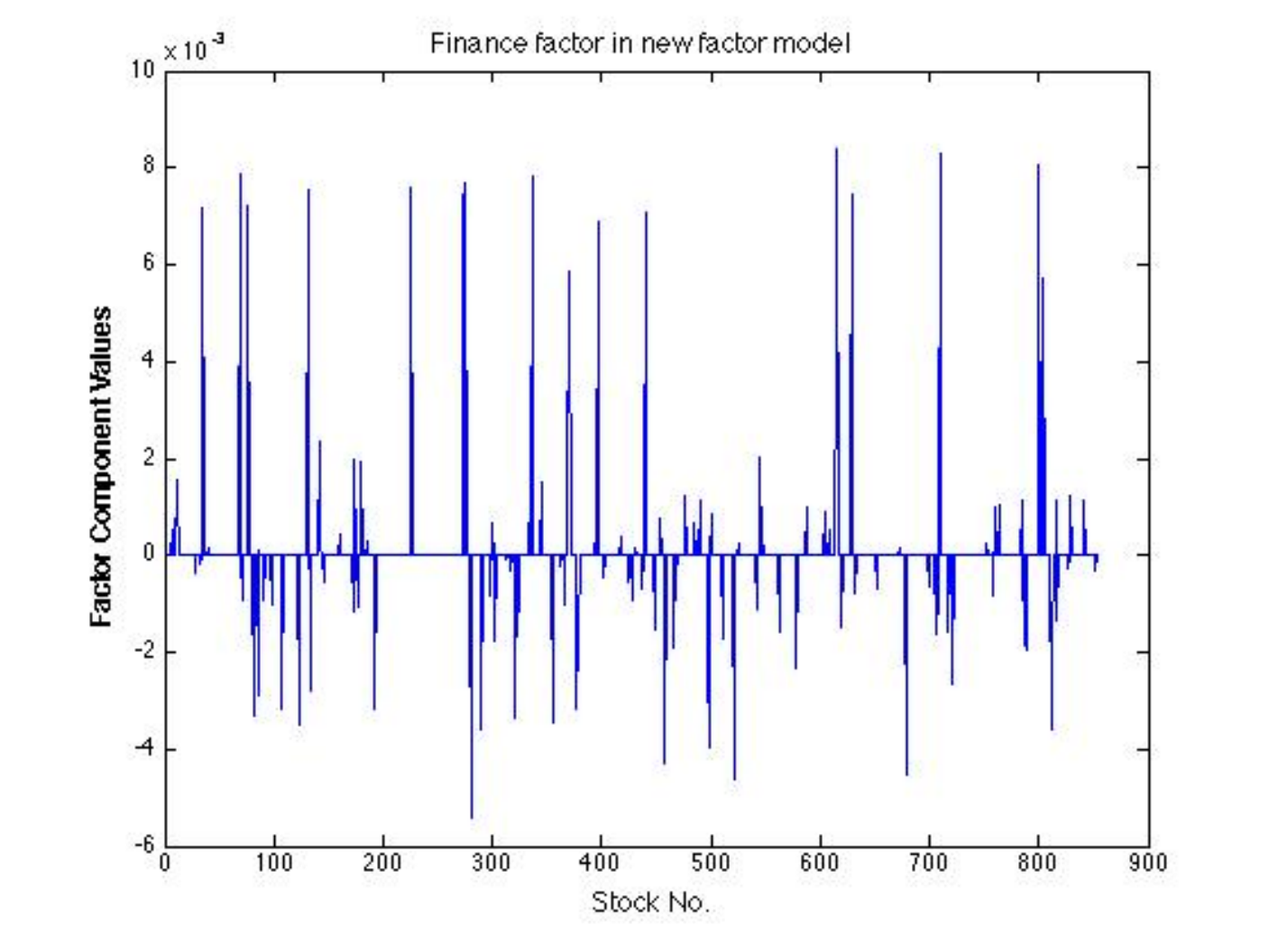}
\end{figure}

The factor models in Figures~\ref{fig41} through \ref{fig45} are computed with
data in the years 1996--2005. We compare the standard factor model with the
new one in terms of the computed factors. Our goal is to demonstrate that the
new factor models indeed can generate more informative factors.

Figure \ref{fig41} depicts one of the factors computed in the standard factor
model. Figure~\ref{fig42} depicts the sector memberships of all the components
that are at least 10\% of the largest component in absolute value. This factor
is dominated by stocks in sectors $1,2$, and $11$, three sectors that are very
different from each other. Since there are as many positive components as
there are negative ones, it is difficult to interpret the factor direction. We
struggle to analyze the graphs of the stock values or to predict any future values.

Figure~\ref{fig43} depicts the factor computed in the new factor model that
corresponds to the technology sector. Most of the sector components are now
negative, indicating the tendency that most technology components move up or
down together over time. This is more consistent with financial intuition
about the phenomenon that stocks in the same sector often move in tandem.
Figure~\ref{fig44} depicts the components in the energy sector. As in
Figure~\ref{fig43}, the energy factor also indicates the tendency to move up
or down together. Instead of the random fluctuations noted in
Figures~\ref{fig41} and \ref{fig42} that don't regard the sectors of the
stocks, we can clearly observe a strong common direction here.

Figure~\ref{fig45} depicts the components in the financial sector. Although
our results supported our hypothesis that the motions of stocks in the same
sector tend to be contingent upon one another, we see that not all sectors
behave the same. In Figure~\ref{fig45}, there seem to be as many positive and
negative components. Due to the nature of finance, we believe stocks in the
financial sector should not all move in the same direction over time.

\begin{figure}[!htbp]
\label{fig46}
\caption{}
\includegraphics[height=3.5in]{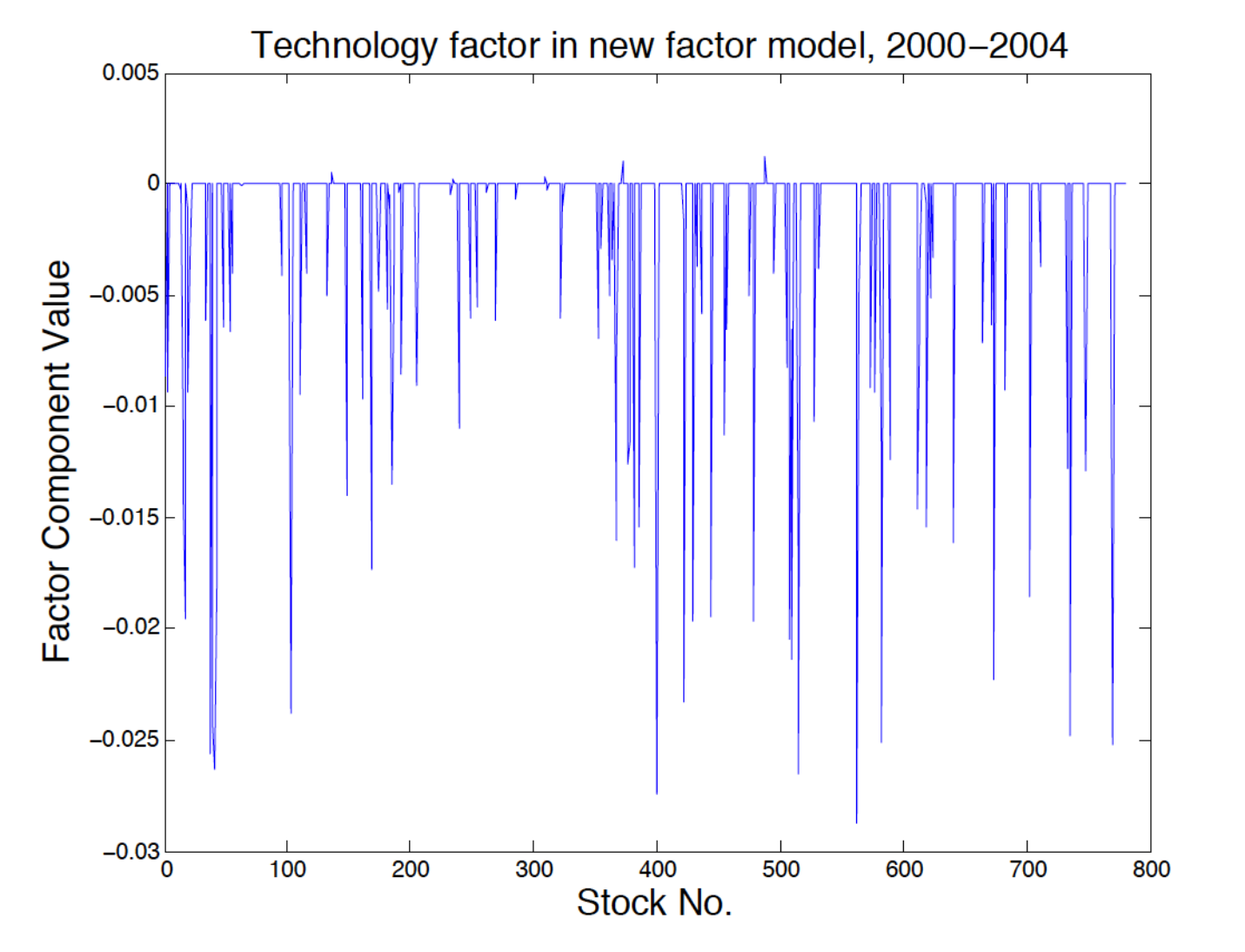}
\end{figure}

Finally, Figure~\ref{fig46} depicts the components in the technology sector
using data during years 2000--2004. While Figure~\ref{fig46} looks quite
different from Figure~\ref{fig43}, indicating large changes in the technology
sector over time, the factor components remain largely negative. This is an
indication that stocks in technology sector consistently move in tandem over
time. We have observed similar behavior in most other sectors.

\section{Illustrations\label{sec4}}

\section{Conclusions and Future Work\label{sec5}}

We sought to understand if taking a sector-based approach to factor analysis
models of the stock market would increase our understanding of stocks and help
us study how different stocks interact. Our experiments with the S\&P 500
index seem to suggest the possibility of additional information in our
computed sector factors. The overwhelming and consistent trends of
Figures~\ref{fig43} and \ref{fig44} do not typically show up in a standard
factor model.

We have only provided a simplified model of reality, though. There are two
major problems with our sectors. Some large companies expand into more than
one sector; Google, for example, is classified as a technology company, but
has recently begun to make phones and cars---respectively, public utilities
and transportation. Additionally, within each sector there are many subgroups.
This is especially an important issue for sectors as large as technology,
which encompasses diverse industries such as computer manufacturing and
microwave devices. medical equipment, physician associations, and medical
training. Future experiments should take these two issues into account.

The goal of any experiment is to apply the results. Our work only focuses on
past data. In order to determine how useful our work in sector-based factor
analysis is, it would be interesting to evaluate the performance of a stock
portfolio where the risk estimation is based on our model in place of a
standard factor model.

Stock will forever remain a mystery; there would be no point in the stock
market if it could be perfectly figured out. We hope, though, that our work
will help unshroud its mysteries just a little more.

\section{Appendix}

\subsection{IBES sector classification\label{appA}}

\begin{verbatim}
01:FINANCE
02:HEALTH CARE
03:CONSUMER NON-DURABLES
04:CONSUMER SERVICES
05:CONSUMER DURABLES
06:ENERGY
07:TRANSPORTATION
08:TECHNOLOGY
09:BASIC INDUSTRIES
10:CAPITAL GOODS
11:PUBLIC UTILITIES
\end{verbatim}

\subsection{EM for factor analysis\label{appB}}

The expected $\log$ likelihood for factor analysis is
\begin{align*}
Q &  =E\left(  \log\left[  \prod\nolimits_{i}(2\pi)^{n}\lvert\Psi\rvert
^{-1/2}\exp\left(  -\frac{1}{2}(X_{i}-\Lambda)^{T}\Psi^{-1}(X_{i}%
-\Lambda)\right)  \right]  \right)  \\
&  =c-\frac{p}{2}\log\lvert\Psi\rvert-\sum\nolimits_{i}\biggl(\frac{1}{2}X_{i}%
^{T}\Psi^{-1}X_{i}-X_{i}^{T}\Psi^{-1}\Lambda E(F\mid X_{i})\\
&  \qquad\qquad\qquad\qquad\qquad\qquad\qquad+\frac{1}{2}\operatorname{tr}[\Lambda^{T}\Psi^{-1}\Lambda
E(FF^{T}\mid X_{i})]\biggr),
\end{align*}
where $c$ is a constant, and $\operatorname{tr}$ is the trace operator.

The last expression is quadratic in $\Lambda(j,\mathcal{I}_{j})$ for each $j$.
This leads to the best optimal estimate for $\Lambda^{\text{new}%
}(j,\mathcal{I}_{j})$. It is also quadratic in each diagonal of $\Psi^{-1}$,
which leads to the best optimal estimate for $\Psi^{\text{new}}$ (see
\cite{10} for more details.)

{\noindent \bf Acknowledgments.} The authors would like to thank
Prof. L.-H. Lim in the Department of Statistics at the University of
Chicago for his invaluable mentorship, guidance, and encouragement
while the authors were working on this project.


\begin{thebibliography}{99} 


\bibitem {1}Bandalos, D.L.; Boehm-Kaufman, M.R. (2008). Four common
misconceptions in exploratory factor analysis". Statistical and Methodological
Myths and Urban Legends: Doctrine, Verity and Fable in the Organizational and
Social Sciences. Taylor \& Francis. pp. 61--87.

\bibitem {2}Bartholomew, D.J.; Steele, F.; Galbraith, J.; Moustaki, I. (2008).
Analysis of Multivariate Social Science Data. Statistics in the Social and
Behavioral Sciences Series (2nd ed.). Taylor \& Francis.

\bibitem {3}Barton, E.S.; Hallbauer, D.K. (1996). "Trace-element and U--Pb
isotope compositions of pyrite types in the Proterozoic Black Reef, Transvaal
Sequence, South Africa: Implications on genesis and age". Chemical Geology
133: 173--199.

\bibitem {4}Bregler, C. and Omohundro, S. M. (1994). Surface learning with
applications to lip-reading. In Cowan, J. D., Tesauro, G., and Alspector, J.,
editors, Advances in Neural Information Processing Systems 6, pages 43-50.
Morgan Kaufman Publishers, San Francisco, CA.

\bibitem {5}Brown, J. D.. "Principal components analysis and exploratory
factor analysis -- Definitions, differences and choices.". Shiken: JALT
Testing \& Evaluation SIG Newsletter. Retrieved 16 April 2012.

\bibitem {6}W. Cheng, Factor Analysis for Stock Performance. Professional
Masters thesis, Worcester Polytechinic Institute, May, 2005.

\bibitem {7}Duda, R. O. and Hart, P. E. (1973). Pattern Classification and
Scene Analysis. Wiley, New York.

\bibitem {8}Everitt, B. S. (1984). An Introduction to Latent Variable Models.
Chapman and Hall, London.

\bibitem {9}Fabrigar et al. (1999). "Evaluating the use of exploratory factor
analysis in psychological research.". Psychological Methods.

\bibitem {10}Ghahramani, Z. and Hinton, G. (1997). The EM Algorithm for
Mixtures of Factor Analyzers. Technical Report CRG-TR-96-1, Department of
Computer Science, University of Toronto.

\bibitem {11}Hochreiter, Sepp; Clevert, Djork-Arn\'{e}; Obermayer, Klaus
(2006). "A new summarization method for affymetrix probe level data".
Bioinformatics 22 (8): 943--9.

\bibitem {12}Ledesma, R.D. and Valero-Mora, P. (2007). "Determining the Number
of Factors to Retain in EFA: An easy-to-use computer program for carrying out
Parallel Analysis". Practical Assessment Research \& Evaluation, 12(2), 1--11

\bibitem {13}Love, D.; Hallbauer, D.K.; Amos, A.; Hranova, R.K. (2004).
"Factor analysis as a tool in groundwater quality management: two southern
African case studies". Physics and Chemistry of the Earth 29: 1135--43.

\bibitem {14}MacCallum, R. (June 1983). "A comparison of factor analysis
programs in SPSS, BMDP, and SAS". Psychometrika 48 (48).

\bibitem {15}Meng, J. (2011). Uncover cooperative gene regulations by
microRNAs and transcription factors in glioblastoma using a nonnegative hybrid
factor model". International Conference on Acoustics, Speech and Signal Processing.

\bibitem {16}Ritter, N. (2012). A comparison of distribution-free and
non-distribution free methods in factor analysis. Paper presented at
Southwestern Educational Research Association (SERA) Conference 2012, New
Orleans, LA.

\bibitem {17}Rubin, D. and Thayer, D. (1982). EM algorithms for ML factor
analysis. Psychometrika, 47(1):69--76.

\bibitem {18}Russell, D.W. (December 2002). "In search of underlying
dimensions: The use (and abuse) of factor analysis in Personality and Social
Psychology Bulletin". Personality and Social Psychology Bulletin 28 (12): 1629--46.

\bibitem {19}SAS Statistics. "Principal Components Analysis". SAS Support Textbook.

\bibitem {20}Schwenk, H. and Milgram, M. (1995). Transformation invariant
autoassociation with application to handwritten character recognition. In
Tesauro, G., Touretzky, D., and Leen, T., editors, Advances in Neural
Information Processing Systems 7, pages 991-998. MIT Press, Cambridge, MA.

\bibitem {21}Spearman, C. (1904). ""General Intelligence," Objectively
Determined and Measured". The American Journal of Psychology 15 (2): 201--292.

\bibitem {22}Sternberg, R.J. (1977). Metaphors of mind: Conceptions of the
nature of intelligence. New York: Cambridge. pp. 85--111.

\bibitem {23}Suhr, D. (2009). "Principal component analysis vs. exploratory
factor analysis". SUGI 30 Proceedings. Retrieved 05 April 2012.

\bibitem {24}Subbarao, C.; Subbarao, N.V.; Chandu, S.N. (December 1996).
"Characterisation of groundwater contamination using factor analysis".
Environmental Geology 28 (4): 175--180.
\end{thebibliography}
\end{document}